\documentclass[reprint,amsmath,amssymb,aps,prl]{revtex4-1}

\usepackage{graphicx}
\usepackage{dcolumn}
\usepackage{bm}
\usepackage{hyperref}
\usepackage{color}
\usepackage[normalem]{ulem}

\begin{document}

\preprint{APS/123-QED}
\title{Periodicity disruption of a model quasi-biennial oscillation}

\author{Antoine Renaud$^1$}
\author{Louis-Philippe Nadeau$^2$}
\author{Antoine Venaille$^1$}
 \email{antoine.venaille@ens-lyon.fr}

\affiliation{
$^1$ Univ Lyon, Ens de Lyon, Univ Claude Bernard, CNRS, Laboratoire de Physique, F-69342 Lyon, France\\
$^2$ Institut des Sciences de la Mer de Rimouski, Universit\'e du Qu\'ebec \`a Rimouski. Rimouski, Qu\'ebec, Canada
}

\date{\today}

\begin{abstract}

The quasi-biennial oscillation (QBO) of equatorial winds on Earth is the clearest example of the spontaneous emergence of a periodic phenomenon in geophysical fluids.In recent years, observations have revealed intriguing disruptions of this regular behaviour, and different QBO-like regimes have been reported in a variety of systems. Here we show that part of the variability in mean flow reversals can be attributed to the intrinsic dynamics of wave-mean flow interactions in stratified fluids. Using a constant-in-time monochromatic wave forcing, bifurcation diagrams are mapped for a hierarchy of simplified models of the QBO, ranging from a  quasilinear model to fully nonlinear simulations.  {The existence of new bifurcations associated with faster and shallower flow reversals, as well as a quasiperiodic route to chaos are reported in these models.} The possibility for periodicity disruptions is investigated by probing the resilience of regular wind reversals to external perturbations.

\end{abstract}

\maketitle

Earth's equatorial stratospheric winds  {oscillate between westerly and easterly mean flow} every 28 months. These low-frequency reversals known as quasi-biennial oscillations are driven by high-frequency waves emitted in the lower part of the atmosphere, and supported by the presence of stable density stratification \cite{baldwin2001quasi}. It is an iconic example of the spontaneous emergence of a periodic phenomenon in a turbulent geophysical flow \cite{vallis2017atmospheric}, with analogues in other planetary stratospheres \cite{dowling2008planetary}, in laboratory experiments \cite{plumb1978instability,semin2018nonlinear}, as well as in idealized numerical simulations \cite{wedi2006direct,couston}. In recent years, increasing attention has been given to the robustness of these regular reversals to external wave forcing and perturbations. Disruptions of this type of oscillations have been observed both in the Earth's atmosphere \cite{osprey2016unexpected,newman2016anomalous} and  in Saturn's atmosphere \cite{fletcher2017disruption}.  In addition, a variety of oscillatory regimes, including non-periodic ones, have been reported in direct numerical simulations of a stratified fluid forced by an oscillating boundary \cite{wedi2006direct} or driven by an explicitly resolved turbulent convective layer \cite{couston}. Non-periodic oscillations have also been reported in global circulation model simulations of the solar interior and Giant planets \cite{rogers2006angular,showman2018atmospheric}.

Until now, the non-periodic nature of the reversals were interpreted as the system's response to transient external variations. For example, the non-periodic disruption of the Earth's QBO and Saturn's QBO-like oscillation have been attributed to the response of equatorial stratospheric dynamics to extratropical perturbations \cite{osprey2016unexpected,newman2016anomalous,fletcher2017disruption}. Also, the existence of nonperiodic regimes in direct numerical simulations of stratified flows has been related to the time variability of the underlying turbulent convective layer \cite{couston,showman2018atmospheric}. Here, we show that the non-periodic nature of the reversals is a fundamental characteristic of stratified fluids by revealing the existence of a vast diversity of oscillatory regimes obtained using a simple steady monochromatic forcing. We further demonstrate that this rich intrinsic variability effectively controls part of the system's response to a transient external variation. Periodicity disruptions are more easily triggered and are increasingly lengthened when the system approaches a bifurcation point.\\

\textbf{Model.} The simplest configuration capturing the dynamics of the quasi-biennial oscillation (QBO, Fig. 1a) is given by a vertical 2D section of a stably stratified Boussinesq fluid, periodic in the zonal (longitudinal) direction, and forced by upward propagating internal gravity waves.
This wave forcing is typically generated by an oscillating bottom boundary meant to represent the effect of tropopause height variations on the stratosphere. The evolution of the horizontally averaged zonal velocity, $\overline{u}$, is governed by a simplified version of the momentum equation
\begin{equation}
\partial_{t}\overline{u}-\nu\partial_{zz}\overline{u}=-\partial_{z}\overline{u^{\prime}w^{\prime}},\label{eq:MeanEvol}
\end{equation} 
where $\nu$ is the kinematic viscosity, $z$ is the upward direction, and $\overline{u^{\prime}w^{\prime}}$ is the Reynolds stress due to velocity fluctuations around the zonal average. In weakly nonlinear regimes, this stress is carried by internal gravity waves, and any process damping the wave amplitude leads to a transfer of momentum from the waves to the mean-flow through the Reynolds stress divergence. Wave properties are also affected by the mean-flow and this interplay results in a complex coupled system. 

To close the dynamical system, one needs to compute the Reynolds stress in (\ref{eq:MeanEvol}).  In this study, the wave field is simulated either by taking into account all nonlinear interactions between waves and mean-flow (hereafter ``nonlinear 2D model") or by considering a simplified closure that neglects wave-wave interactions, together with a WKB approach \cite{plumb1977interaction}. The latter approach (hereafter ``quasilinear 1D model") has proven to be successful in explaining the spontaneous emergence of low-frequency periodic flow reversals \cite{plumb1977interaction,vallis2017atmospheric} and synchronisation with an external forcing \cite{rajendran2016synchronisation}. By assuming horizontally averaged dynamics, this quasilinear model is much simpler than the original flow equations but has nevertheless a large number of degrees of freedom since an infinite number of vertical oscillatory modes are possible. The 1D model is thus a natural starting point to investigate how periodic reversals are destabilized when the forcing strength is increased.

\begin{figure*}
\centering\includegraphics[width=0.8\textwidth]{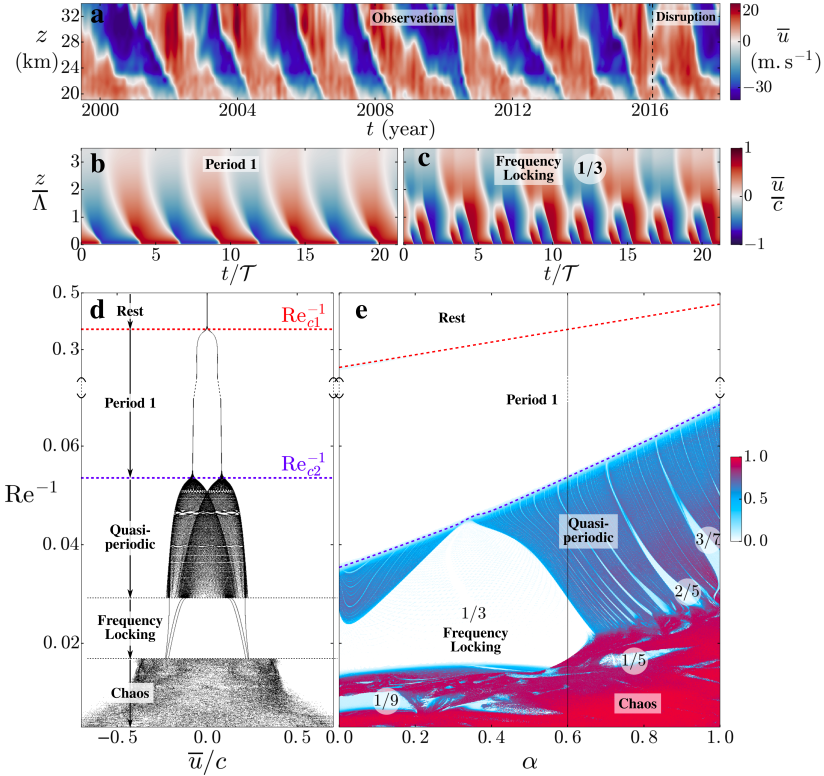}
\caption{\label{fig:Plumb_2Ddiag}\textbf{Bifurcations in the 1D quasilinear model.} \textbf{a.} Observations of the Earth's atmospheric quasi\--biennial oscillations. Hovm\"oller diagram of monthly averaged zonal winds measured by radiosonde in the lower stratosphere above Singapore ($1.4^{\circ}\mathrm{N}$) \cite{QBODATA}.  \textbf{b.} Hovm\"{o}ller diagram of the mean-flow $\overline{u}(z,t)$ in stationary regime for $\alpha=0.6$ and $\mathrm{Re}^{-1}=0.06$, using the 1D quasilinear model. \textbf{c.} Same as b, but using  $\mathrm{Re}^{-1}=0.025$. See Supplemental Material for similar time-height plots in other regimes. \textbf{d.} Bifurcation diagram for $\alpha=0.6$, obtained for each value of $\mathrm{Re}^{-1}$ by considering the value of $u$ at two different heights $z_1$ and $z_2$, and then by plotting $u(z_2)$ at times when $u_(z_1)$. The dashed red line corresponds to the first bifurcation (from rest to period-1) occurring at $\mathrm{Re}_{c1}$. The dashed blue line corresponds to the second bifurcation (from period-1 to quasiperiodic) occurring at $\mathrm{Re}_{c2}$. \textbf{e.} Bifurcation diagram in parameter space $\left(\alpha,\mathrm{Re}^{-1}\right)$. The coloured field is an empirical estimate of the area of the attractor projected in panel d.  Low values (in white) correspond to QBO-like regions (period-1) and frequency locked regions (rational numbers). The vertical black line at $\alpha=0.6$ corresponds to the bifurcation diagram plotted in panel c.}
\end{figure*}

We consider a standing wave pattern with wavenumber $k$ and frequency $\omega$, forcing a stratified fluid with buoyancy frequency $N$, for which the background stratification is maintained by Newtonian cooling with damping rate $\gamma$. Together, the Newtonian cooling $\gamma$, and the viscosity $\nu$, damp the wave amplitude over a characteristic e-folding length $\Lambda=\alpha k c^{4} /(\nu N^3)$, where $c=\omega/k$ is the zonal phase speed, and where $\alpha=\nu N^2/(\nu N^2 +\gamma c^2)$ is the ratio of viscosity to Newtonian cooling in wave damping. Another essential parameter of the problem is the effective Reynolds number  $\mathrm{Re}=\mathcal{F}_{0} \Lambda/\left(c \nu \right)$, where $\mathcal{F}_{0}={\left(u^{\prime}_0w^{\prime}_0\right)}_{\mathrm{r.m.s}.}$ is the wave forcing strength at the bottom boundary. This wave forcing strength further sets a characteristic time scale of low-frequency flow reversals $\mathcal{T}=c\Lambda/\mathcal{F}_{0}$ \cite{vallis2017atmospheric}. Supplemental Material provides details on the simulations as well as estimates of the key parameters of the the quasilinear and nonlinear models, as well as for the Earth's stratosphere. 

The parameter range used in the simulations is close to that used in the pioneering work on the subject \cite{holton1972updated,plumb1977interaction}, and presented in standard textbooks on geophysical fluid dynamics  \cite{vallis2017atmospheric}. Notice that the effective Reynolds number is based on an eddy viscosity, meant to represent the  turbulent eddy motion at scales smaller than the internal gravity waves and used as a subgrid-scale parameterization for turbulence in coarse-grained climate models. The actual Reynolds number of the atmosphere based on the kinematic viscosity of the air is higher by many orders of magnitude than the effective Reynolds number. A self-consistent theory for the QBO would require to infer the eddy viscosity from the knowledge of the actual Reynolds number and other problem parameters, but this conundrum has up to now be out of reach. Here we follow a common practice in  geophysical fluid dynamics that amounts to: (i) use an eddy viscosity to describe bifurcations occurring under an increase in forcing amplitude, and (ii) test the robustness of these bifurcations in more complex members  of the hierarchy of geophysical flow models \cite{dijkstra2013nonlinear}. 
\\

\textbf{Bifurcation diagrams.} To map the bifurcation diagram of the quasilinear 1D model, we performed a large number of simulations spanning effective Reynolds numbers between $\mathrm{Re}=2$ and $330$, covering roughly the relevant range for the Earth's stratosphere (Table 1). For sufficiently low values of $\mathrm{Re}$, the system has only one attractor: a stable point at $\overline{u}=0$. A first bifurcation occurs above the  critical value $\mathrm{Re}_{c1}\approx 4.25/(1+\alpha)$ \cite{yoden1988new}, for which the zonally averaged velocities are attracted towards a limit cycle~\cite{plumb1977interaction,yoden1988new} corresponding to horizontal mean-flow reversals and downward phase propagation (Fig. 1b). This period-1 cycle arguably reproduces the salient features of the observed QBO  before the disruption event of 2016 (Fig. 1a). Figure 1d shows a bifurcation diagram plotted for increasing Reynolds numbers. A second bifurcation from periodic to quasi-periodic regimes occurs above the critical value $\mathrm{Re}_{c2}$. Additional bifurcations occur at higher Reynolds numbers, with transitions to frequency-locked regimes, and chaotic regimes. The term 'frequency-locking' is often used where a nonlinear oscillator forced at some frequency exhibits, as a dominant response, an oscillation at the forcing frequency. By extension, we use this term here to describe synchronisation between oscillating modes of the dynamical system. As $\mathrm{Re}$ increases, new oscillating modes appear in the vertical structure of the mean flow. For example, a unique frequency is observed at all heights for the period-1 limit cycle shown in Fig. 1b, while faster reversals are observed in the lower levels for the frequency-locked regime shown in Fig. 1c. Ultimately, in chaotic regimes, the superposition of these modes yields a fractal-like structure of nested flow reversals (Fig. S1 in Supplemental Material). Such regimes with faster reversals in the lower layers have also been reported in direct numerical simulations driven by a convective boundary layer \cite{couston,showman2018atmospheric}.

The quasiperiodic regime occurring at $\mathrm{Re}>\mathrm{Re}_{c2}$ is embedded with a complicated set of frequency-locked regimes (Fig. 1d).  The global structure of the bifurcation diagrams is better appreciated by considering, in Fig. 1e, the two-dimensional parameter space spanned by the Reynolds number $Re$ and the parameter $\alpha$. This figure shows a range of parameters where frequency locked regions are organized into a sequence of staircases, qualitatively similar to Arnold's tongues \cite{arnold1961small}.

Transition to chaos in a similar 1D quasilinear model was reported in Ref. \cite{kim2001gravity}, which focused only on the purely viscous case, $\alpha=1$, with other boundary conditions relevant for the solar tachocline. In fact, this behavior occurs generically in nonlinear systems, with numerous examples in hydrodynamics  \cite{swinney1983observations}.

In the case of internal gravity wave streaming, frequency locked states organized into Arnold's tongues were found when the 1D quasilinear model is coupled to an external low frequency forcing mimicking seasonal forcing \cite{read2015}, which is reminiscent of synchronisations phenomena  {in models} of El Nino Southern Oscillations \cite{tziperman1994nino,jin1994nino}. Here, by considering a simple monochromatic forcing, and by covering the full parameter space $Re-\alpha$, we bring to light an unforeseen intrinsic dynamical structure of the underlying quasilinear model.

\begin{figure}
\centering
\includegraphics[width=\columnwidth]{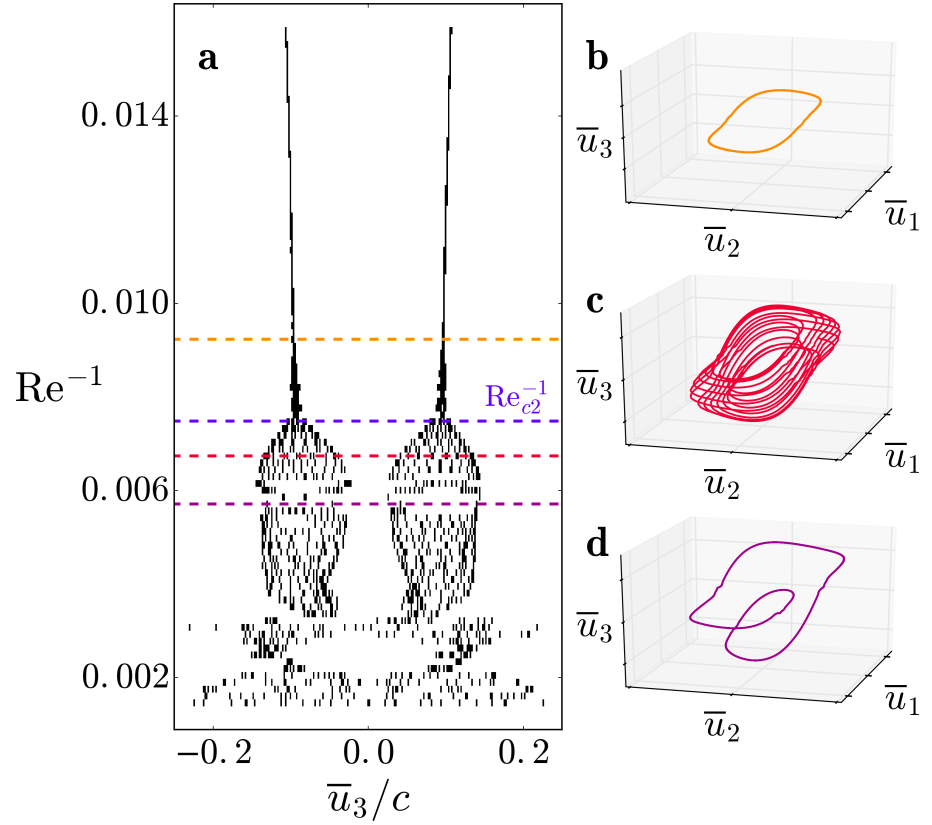}
\caption{\label{fig:BifDiag_GCM}\textbf{Bifurcations in Navier-Stokes simulations. a.} 
As in Fig 1d, but showing the bifurcation diagram obtained with the nonlinear simulations, using $\alpha=0.6$. 
Selected values of $\mathrm{Re}$ (marked in orange, red and purple) correspond to panels b, c, and d. The dashed blue line corresponds to the second bifurcation (from period-1 to quasiperiodic) occuring at $\mathrm{Re}_{c2}$. \textbf{b.} Phase space trajectory for  $\mathrm{Re}^{-1}=0.009$, projected on a 3D space defined by velocities at three different heights: $(\overline{u}_{1},\overline{u}_{2},\overline{u}_{3})=(\overline{u}(z=0.5\Lambda,t),\overline{u}(z=1.5\Lambda,t),(z=3\Lambda,t))$. \textbf{c.} Same as b, but using $\mathrm{Re}^{-1}=0.0068$. \textbf{d.} Same as b, but using $\mathrm{Re}^{-1}=0.0058$.}
\end{figure}

The 1D quasilinear model is a highly truncated version of the original flow equations. It is thus crucial to see whether the aforementioned bifurcations occur in Navier-Stokes simulations of the fully nonlinear dynamics, including both wave-mean and wave-wave interactions. In Fig. 2a, we performed more than 200 two-dimensional numerical simulations to build a diagram similar to the one obtained with the 1D quasilinear model. These simulations show that the route to chaos is robust to the presence of nonlinear interactions between waves and mean-flow, with transitions from periodic solutions (Fig. 2b)  to quasiperiodicity (Fig. 2c), to frequency locking (Fig. 2d) and eventually to chaos. However, significant differences from the quasilinear case are observed in the nonlinear simulations, where bifurcations occur at different effective Reynolds numbers, and where new dynamical regimes emerge. For instance, the large region of period-3 frequency locking obtained in the 1D quasilinear model (Fig. 1c) is replaced by a thin region of period-2 frequency locking for which the symmetry $U \rightarrow -U$ is broken (Figs. 2a and 2d).\\

\textbf{Response to external perturbations.} By considering a fixed monochromatic wave forcing, we show above that quasi-periodicity arises naturally at steady state in the stratified fluid. This fixed forcing contrasts however with the actual QBO signal, which is driven by time-varying wave forcing and extra-tropical perturbations. In the following, we investigate how the presence of a bifurcation point influences the resilience of a given period-1 QBO-like oscillation to external variability by considering the effect of a time-dependent perturbation superimposed on its reference monochromatic wave forcing.

We first consider the effect of a time-dependent pulse in wave forcing strength, $\mathcal{F}_{0}$, mimicking the reported sudden increase in wave activity at the equator in the winter preceding the observed periodicity disruption of 2016 (see Supplemental Material for details on the perturbation). From a dynamical point of view, this perturbation suddenly drives the system out of its limit cycle, until it eventually relaxes back to its original period-1 oscillation over a characteristic time $\tau$. Figures 3a and 3b show examples of transient recovery periods for two values of the effective Reynolds number using the nonlinear model. In each case, the time evolution of the mean-flow displays short eastward-flow structures sandwiched between broader westward wind patterns.  {These higher vertical modes of oscillations, frequently excited in transient disrupted regimes, share qualitative similarities with the periodicity disruption observed in 2016. }  

Figure 3d shows that the characteristic timescale for recovery diverges as the system approaches the bifurcation point $\mathrm{Re}_{c2}$. We found similar responses to a pulse in zonal mean momentum, and for the spin-up of the system from a state of rest. The recovery time's divergence is observed both in the quasilinear model and the nonlinear model. The swift increase in recovery timescale observed as the system approaches a bifurcation point is  {a generic feature of dynamical systems} often referred to as ``critical slowing down" \cite{scheffer2009early,kuehn2011mathematical}. In the climate system context, critical slowing down has proven to be useful in detecting early warnings of a bifurcation point~\cite{lenton2011early}. \\

\begin{figure*}
\centering\includegraphics[width=\textwidth]{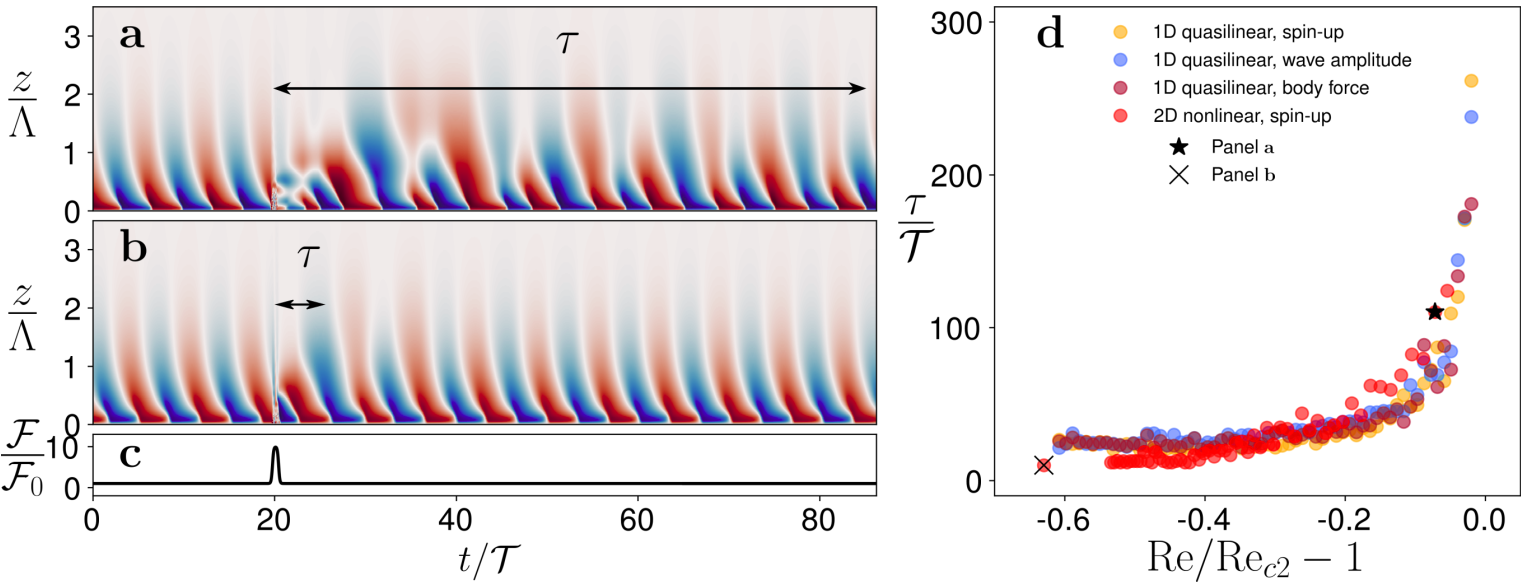}
\caption{\label{fig:qboBUMPOBS}\textbf{Critical slowing down. }  \textbf{a.} Hovm\"oller diagrams of the mean-flow, $\overline{u}(z,t)$, in the Navier-Stokes simulations, perturbed by a pulse in wave forcing for $\mathrm{Re}=125$. \textbf{b.} Same as a, but using $\mathrm{Re}=50$. \textbf{c.} 
Time evolution of the wave forcing strength $\mathcal{F}\left(t\right)$. \textbf{d.} Characteristic recovery timescale as a function of relative distance to the second bifurcation point, $1-\mathrm{Re}/\mathrm{Re}_{c2}$ for different forcing procedures decribed in Supplemental Material. The star and the cross correspond to the two points associated with the nonlinear computations of panels a and b, respectively. We used $\mathrm{Re}_{c2}=24.3$ for the 1D quasilinear model and $\mathrm{Re}_{c2}=135$ for the nonlinear 2D model.}
\end{figure*}

\textbf{Conclusions and perspectives.} Our study demonstrates that erratic mean flow reversals are recovered with a simple monochromatic wave-forcing, provided that the forcing strength is sufficiently large. This suggests that similar states previously observed  with more complex forcing \cite{couston,showman2018atmospheric} can partly be attributed to the intrinsic dynamics of stably stratified fluids, rather than to fluctuations of the forcing itself. The quasiperiodic route to chaos found  both in our  quasilinear and fully nonlinear simulations reveal that increasing the forcing strength leads to the excitation of fast and shallow bottom-trapped modes nested in deeper and slower vertical modes. These fast bottom-trapped reversals are also excited  during the transient response to an external perturbation. Most important, our results have crucial implications for the interpretation of the variability of a QBO-like oscillation: 
(i) the existence of a second bifurcation is robust in a hierarchy of models (suggesting that it may exist for the atmosphere), 
(ii) the proximity to this second bifurcation has a strong effect on the response of the oscillation to external perturbations, and consequently
(iii) the intrinsic variability of a given oscillation is key to interpret its response to external perturbations.

Several aspects of actual planetary flow such as seasonal forcing, rotation, meridional circulation and two-way  coupling between stratospheric and tropospheric dynamics are omitted in the simplified flow models considered in this letter. The interplay between intrinsic modes of variability and these additional features will need to be addressed in future work, but we expect that the the existence of a second bifurcation as well as the critical slowing down approaching this bifurcation will be robust through the whole hierarchy of geophysical flow models with stable stratification. Exploratory 3D simulations with rotation \-- presented in supplementary materials \-- comfort our 2D results.


\pagebreak
\newpage
\setcounter{equation}{0}
\setcounter{figure}{0}
\setcounter{table}{0}
\setcounter{page}{1}
\makeatletter
\renewcommand{\theequation}{S\arabic{equation}}
\renewcommand{\thefigure}{S\arabic{figure}}
\renewcommand{\bibnumfmt}[1]{[S#1]}
\renewcommand{\citenumfont}[1]{S#1}

\begin{widetext}
\begin{center}
    \textbf{\large Supplemental Material for `Periodicity disruption of a model quasi-biennial oscillation'}\\[15pt]
    Antoine Renaud$^1$, Louis-Philippe Nadeau$^2$, Antoine Venaille$^{1}$\\[3pt]
    \emph{$^1$ Univ Lyon, Ens de Lyon, Univ Claude Bernard,\\
    CNRS, Laboratoire de Physique, F-69342 Lyon, France\\
$^2$ Institut des Sciences de la Mer de Rimouski, Universit\'e du Qu\'ebec \`a Rimouski. Rimouski, Qu\'ebec, Canada}\\
    (Dated: \today )
\end{center}

\section*{Methods and details on numerical simulations}

\textbf{1D quasilinear simulations.} Using a static Wentzel-Kramers-Brillouin (WKB) approximation to compute the wave field for a given mean-flow \cite{Splumb1978instability}, the wave-induced Reynolds stress $\overline{u^{\prime}w^{\prime}}$ is parametrised by the formula given in Eq. (\ref{eq:Parametrisation}).

\begin{equation}
\overline{u^{\prime}w^{\prime}}\left(z\right)=\sum_{i=1}^{2} (-1)^i\mathcal{F}_{0}\exp\left\{-\frac{1}{\Lambda}\int_{0}^{z}\mathrm{d}z^{\prime}\left(\frac{\alpha}{\left(1-(-1)^i\overline{u}\left(z^{\prime}\right)/c\right)^{4}}+\frac{1-\alpha}{\left(1-(-1)^i\overline{u}\left(z^{\prime}\right)/c\right)^{2}}\right)\right\}\label{eq:Parametrisation}.
\end{equation}

This formula is derived under the hydrostatic balance assumption (valid in the limit $k|c\pm\overline{u}|/N\rightarrow 0$) and the weak damping assumption (valid in the limit $\gamma/(k|c\pm \overline{u}|)\ll1$ and $\nu N^{2}/(k|c\pm\overline{u}|^{3})\ll 1$). Assuming that the characteristic vertical length for $\overline{u}$ is $\Lambda$, then the small parameter needed in the WKB approach is the Froude number $Fr = c/(\Lambda N)\rightarrow 0$. In practice, the different assumptions are most certainly violated. However, this set of equations has long been recognized as a useful model to probe the salient features of QBO reversals.

We solve numerically using a centered second-order finite difference method with grid size $\delta_{z}=H/60$, and a second-order Adams-Bashforth scheme with time-step $\delta_{t}=0.005\mathcal{T}$; $\mathcal{T}=c\Lambda/\mathcal{F}_{0}$. 
A no-slip condition is used at the bottom boundary, $z=0$, and a free-slip condition is used at the upper boundary, $z=H$. 
Singularities in Eq. (\ref{eq:Parametrisation}) appear when $\overline{u}=c$ (critical layers). These singularities are treated as follows: at a given height $z=z_c$, if the absolute value of $\overline{u}$ reaches locally a value higher than $c$, then the corresponding exponential is set to zero for all $z\geq z_{c}$.
The definition and value of each of the model's dimensionless numbers are given in table \ref{tab:Param}.\\

\textbf{2D nonlinear simulations.} The fully nonlinear simulations are conducted using the MIT general circulation model  \cite{SMarshall1997} solving the 2D Navier\--Stokes equations under the Boussinesq and hydrostatic approximations
\begin{equation}
    \begin{cases}
        \partial_{t}u+\mathbf{u}\cdot\nabla u & =-\partial_{x} \phi+\nu\nabla^{2}u-\gamma_u \left(u-u_0\right) \\
        0 &  = -\partial_{z}\phi+b\\
\partial_{t}b+\mathbf{u}\cdot\nabla b&= \kappa\nabla^{2} b-\gamma\left(b-b_0\right)\\
        \nabla\cdot\mathbf{u}&=0
    \end{cases},\label{eq:NumModel}
\end{equation}
where $\mathbf{u} = u {\hat{\textbf{\i}}}+ w {\hat{\textbf{k}}} $
is the velocity field, $\mathbf{u}_{\rm h}$ is its projection on the horizontal plane $(x,y)$; $b=g\left(\rho_{0}-\rho\right)/\rho_{0}$ is the buoyancy; $\rho$ is the density and $\rho_{0}$ is a reference density; $g$ is the gravitational acceleration; $\phi=P/\rho_{0}+gz$; $P$ is the pressure; $\nu$ is the viscosity coefficient; $\kappa$ is the buoyancy diffusion coefficient; $\gamma_u$ and $\gamma$ are the rates at which the momentum and buoyancy are linearly restored to the reference profiles $u_0$ and $b_0$, respectively.

The domain is a Cartesian grid, periodic in the zonal direction, with zonal length $L_{x}=2\pi/k$ and height $H$ . The horizontal and vertical resolutions are respectively $\delta_x=L/26$ and $\delta_{z}=H/200$. A free-slip condition is used at the bottom boundary, while a free-surface condition is used at the top.

The zonal momentum equation is forced at the bottom boundary using a linear velocity relaxation  $\gamma_u=\delta_b/\tau_{u}$, where $\tau_{u}$ is a relaxation timescale, and $\delta_b$ is a delta function equal to $1$ for the bottom grid-point and $0$ for all other vertical levels. In this last grid-point, velocity is relaxed to a zonally periodic standing wave pattern 
\begin{equation}
u_0=\sqrt{\frac{4N\mathcal{F}_0}{\omega}}\cos\left(kx\right)\cos\left(\omega t\right), \label{eq:bottom_dns}
\end{equation}
where $\mathcal{F}_0$ controls the wave momentum flux amplitude at the bottom. This forcing is thought to generate a standing internal gravity wave field while enforcing an effective no-slip condition for the mean flow $\overline{u}$. Buoyancy is relaxed to the linear profile $b_0=N^{2}z$. To avoid any wave reflection at the upper free-surface, the vertical grid spacing and the Newtonian cooling are both increased in the $20$ upper grid layers.\\

\textbf{Poincar\'e sections.} For each combination of parameters $\left(\mathrm{Re},\alpha\right)$, experiments are first spun-up over a time $t_{\rm e}=1500\mathcal{T}$ where $\mathcal{T}=c\Lambda/\mathcal{F}_{0}$. This time is sufficient for the system to reach its attractor. To combine the information of more than $10^{6}$ simulations into a single bifurcation diagram, we first select two vertical levels: $z_1$ near the surface and $z_2$ aloft. 
Resuming the simulation at statistical equilibrium ($t>t_{\rm e}$), we store the values of $\overline{u}(z_2)$ that intersects $\overline{u}(z_1)=0$ in the set
\begin{equation}
    \mathbb{O}_{\mathrm{Re},\alpha}=\left\{\;\overline{u}\left(z_{2},t\right)\left|\;\overline{u}\left(z_{1},t\right)=0\right.\right\}.\label{eq:Mat_PS}
\end{equation}
The simulations are stopped once $200$ values are stored (i.e. after $200$ reversals of the lower-level mean-flow $\overline{u}\left(z_{1}\right)$). For each simulation associated with couples of parameters $(\mathrm{Re},\alpha)$, we build an histogram of the values stored in (\ref{eq:Mat_PS}), using $1000$ bins in the range $\left[-c,c\right]$. Histograms corresponding to all values of $\mathrm{Re}$ for a fixed $\alpha=0.6$ are drawn horizontally in figure 1c using a binary colour-map. 
\begin{table*}
\centering
\begin{tabular}{|c l|c|c|c|}
    \cline{3-5}
    \multicolumn{2}{c|}{}& 1D quasilinear simulations & 2D Nonlinear simulations & Stratosphere\\
    \hline
    $\mathrm{Re}$&$=F_{0}\Lambda/\left(c\nu\right)$&$15-350$&$50-600$&$2-400$ \\
    \hline
    $\alpha$ & $=\nu N^{2}/\left(\nu N^{2}+\gamma c^{2}\right)$ &$0-1$& $0.25$ &$0-0.3$\\
    \hline
    $Pr$&$=\nu/\kappa$&$\infty$&$740$&  N.A.\\
      \hline
       $\mathrm{Fr}$&$=c / \left(\Lambda N\right)$&$ \mathrm{Fr} \to 0 $&$ 0.06$&$0.1$\\
    \hline
     \  & ${\color{white} =}\omega/N$&$ \omega/N \to 0$&$0.1$&$10^{-4}-10^{-3}$\\
        \hline 
    \  & ${\color{white} =}\omega\tau_{u}$&$0$&$0.1$&N.A.\\
    \hline
     \  & ${\color{white} =}H/\Lambda$&$3.5$&$4.1$&$1.5$\\  
    \hline
  \end{tabular}
  \caption{\label{tab:Param}Dimensionless parameters. For the stratospheric values, we considered $F_{0}= 3-10\times 10^{-3}\,\mathrm{m}^{2}\mathrm{s}^{-2}$, $c=25\,\mathrm{m}\mathrm{s}^{-1}$, $\gamma=0.5-1.5\times 10^{-6}\,\mathrm{s}^{-1}$ and $N=2.2\times 10^{-2}\,\mathrm{s}^{-1}$ based on \cite{Svallis2017atmospheric}. We considered $\Lambda\sim 10\,\mathrm{km}$ based on the QBO observations (see fig. 1a in the letter). We consider  $\nu=0.01-0.3\,\mathrm{m}^{2}\mathrm{s}^{-1}$ corresponding to turbulent vertical eddy-diffusivity measured in the lower stratosphere (see e.g. \cite{SHaynes2005,Sholton1972updated}). The closure used in the 1D quasilinear model  reduces the number of dimensionless parameters down to three by assuming $\omega/N \rightarrow 0$ (hydrostatic approximation) and a low  Froude number limit $Fr\rightarrow 0$ (WKB approximation).}
\end{table*}

To collapse the information of the Poincar\'e sections into a 2D bifurcation diagram $(\alpha, \mathrm{Re}^{-1})$, we compute the ratio of populated bins to the total number of reversals for each histogram in the set (\ref{eq:Mat_PS}). This ratio with values in $\left[0,1\right]$ provides an empirical estimate of each histogram's distribution and allows for an extensive classification of the different dynamical regimes (see Fig. 1d in the letter)\\

\textbf{Recovery from a perturbation.}  We consider perturbations to a given period-1 QBO-like oscillation.
Three types of external perturbations are considered at $t=t_p$: (i) a pulse in wave amplitude (representing a sudden increase of the underlying tropical convection) (ii) a body force acting directly on the mean flow (representing a reorganization of the mean flow due to extratropical perturbations) (iii) a reboot of the oscillations from a state of rest (the recovery time is then equivalent to the spin-up time).

Perturbation (i) is modeled using a time-dependent momentum flux amplitude $\mathcal{F}_0$ in Eq. (\ref{eq:Parametrisation}) for the quasilinear model and in Eq. (\ref{eq:bottom_dns}) for the nonlinear simulations:
\begin{equation}
\mathcal{F}_0\left(t\right)=\mathcal{F}_{0,p}\left(1+9e^{-\frac{1}{2}\left(10\frac{t-t_{p}}{T_{\rm qbo}}\right)^{10}}\right),
\end{equation}
where $\mathcal{F}_{0,p}$ is a constant forcing amplitude corresponding to a periodic regime with  period $T_{\rm qbo}$. Perturbation (ii) is represented in the quasilinear model by an additional body forcing term  $\mathcal{F}_{\mathrm{bulk}}$ in the r.h.s. of the mean flow equation (1):
\begin{equation}
\mathcal{F}_{\mathrm{bulk}}\left(z,t\right)=
\frac{10}{T_{\rm qbo}} e^{-\frac{1}{2}\left(10\frac{t-t_{p}}{T_{\rm qbo}}\right)^{10}} e^{-800\left(\frac{z-z_{p}}{z_{\rm max}}\right)^{2}},
\end{equation}
where $z_{p}=0.2z_{\rm max}$ sets the height. Results are insensitive to the specific choices of $z_{p}$.

For all three types of external perturbations, the system is driven away from its steady state period-1 limit cycle and then freely recovers back to the cycle. To estimate the recovery timescale, we first introduce the running mean-square
\begin{equation}
\left\langle\overline{u}^{2}\right\rangle\left(t\right)=\frac{1}{T_{\rm qbo}H}\int_{t-T_{\rm qbo}/2}^{t+T_{\rm qbo}/2}\,\int_{0}^{H}\,\overline{u}^{2}\left(z,t^{\prime}\right)\mathrm{d}z\mathrm{d}t^{\prime},
\end{equation}
where $T_{\rm qbo}$ is the period of the limit cycle. At steady state equilibrium, this running mean-square has a constant value $\langle\overline{u}^{2}\rangle_{\infty}$. 
Assuming a pulse shorter than the period of the limit cycle (see Fig. 3c in the letter), occurring at time $t_{p}$, the recovery timescale is then defined by
\begin{equation}
\label{recovT}
\tau=\min_{\Delta t\geq 0} \left\{ \frac{\langle\overline{u}^{2}\rangle\left(t_{p}+\Delta t\right)-\langle\overline{u}^{2}\rangle_{\infty}}{\langle\overline{u}^{2}\rangle_{\infty}}\leq 0.2 \right\}. 
\end{equation}

We reproduced figure 3d of the letter in logarithmic scale in order to exhibit the power-law like scaling of the recovery timescale as the system approaches $\mathrm{Re}_{c2}$. It proved very difficult to deduce a precise value for the critical exponent as the uncertainty on the value of $\mathrm{Re}_{c2}$ echoes on it. However, the critical exponent remains close to $-1$.

\begin{figure}
\centering\includegraphics[width=0.4\linewidth]{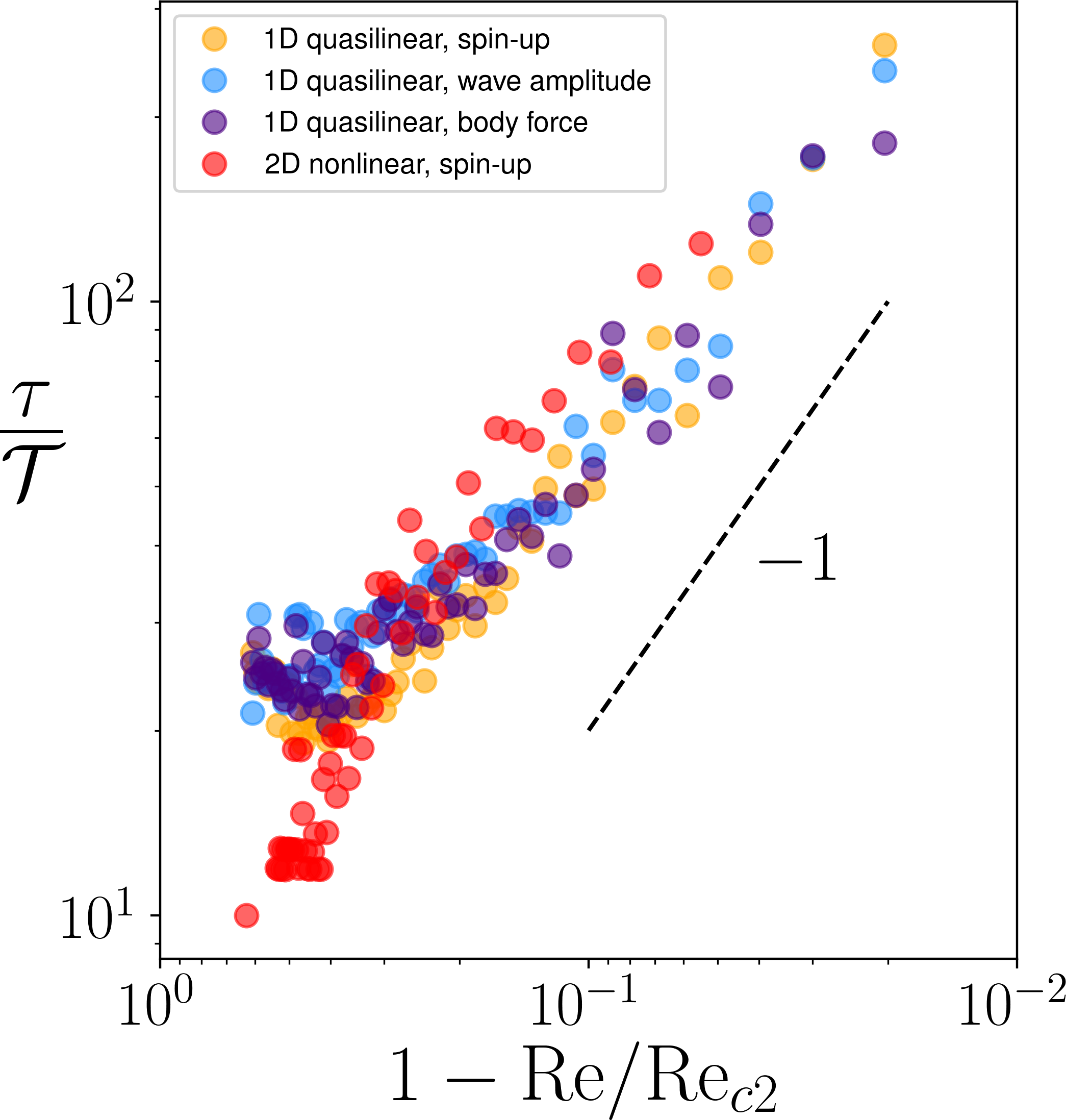}
\caption{\label{fig:BifDiag_Plumb_S6}\textbf{Spin-up and recovery times in log-scale.} Characteristic recovery and spin-up timescales as a function of the relative distance to the second bifurcation point, $1-\mathrm{Re}/\mathrm{Re}_{c2}$, represented in log-scale with revert $x$-axis. The orange markers correspond to the spin-up time for the 1D quasi-linear model. The blue and purple markers are obtained with the quasilinear model, and correspond to the recovery from a pulse in the wave amplitude and from a direct body force, respectively.  The red markers correspond to the spin-up time with the 2D nonlinear model. We used $\mathrm{Re}_{c2}=24.3$ for the 1D quasilinear model and $\mathrm{Re}_{c2}=135$ for the 2D nonlinear model. A dashed line representing a power law with exponent $-1$ is provided to guide the reader.}
\end{figure}

\section*{Vertical flow  structure in different regimes and effect of resolution}
\begin{figure}
\centering\includegraphics[width=0.8\linewidth]{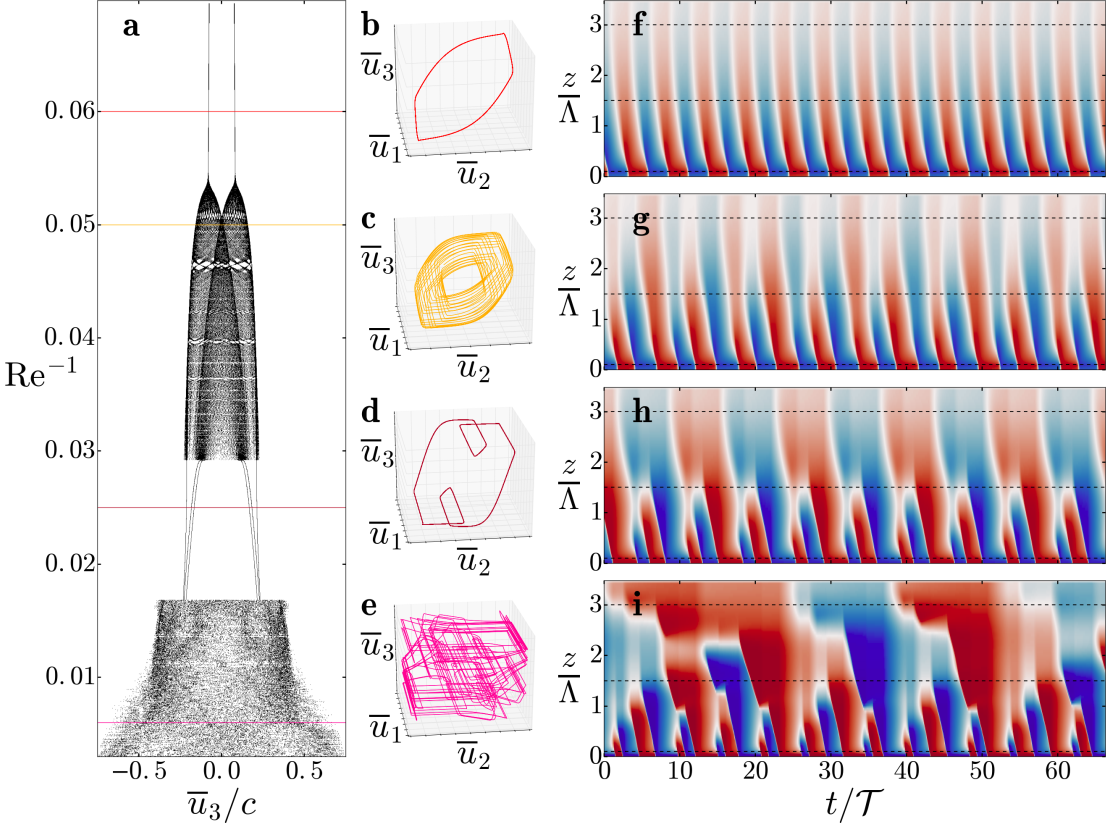}
\caption{\label{fig:BifDiag_Plumb_S1}\textbf{Bifurcations in the 1D quasilinear model. a.} A Poincar\'{e} section is shown for varying values of $\mathrm{Re}^{-1}$ and $\alpha=0.6$ (see Methods). 
\textbf{b.} Projection of phase-space trajectory in a 3D space $(\overline{u}_{1},\overline{u}_{2},\overline{u}_{3})=(\overline{u}(z=0.1\Lambda),\overline{u}(z=1.5\Lambda,t),\overline{u}(z=3\Lambda,t))$ for $\mathrm{Re}^{-1}=0.059$. \textbf{c.} Same for $\mathrm{Re}^{-1}=0.045$. \textbf{d.} Same for $\mathrm{Re}^{-1}=0.025$. \textbf{e.} Same for $Re^{-1}=250$. \textbf{f.} Hovm\"oller diagram of the mean-flow $\overline{u}\left(z,t\right)$ for $\mathrm{Re}^{-1}=0.059$. Time is rescaled by $\mathcal{T}=c\Lambda/\mathcal{F}_{0}$. The velocity $\overline{u}$ ranges from to $-c$ (blue) to $+c$ (red). The horizontal dotted lines highlight the height $z=0.1\Lambda$, $z=1.5\Lambda$ and $z=3\Lambda$, associated with the 3D projections plotted in panels b to e.  \textbf{g.} Same for $\mathrm{Re}^{-1}=0.045$. \textbf{h.} Same for $\mathrm{Re}^{-1}=0.025$. \textbf{i.} Same for $\mathrm{Re}^{-1}=0.0004$.}
\end{figure}

\textbf{Bifurcations in the quasilinear model.} In order to develop intuition for the underlying dynamics of the bifurcation diagrams of Fig. 1d, we show in Fig. \ref{fig:BifDiag_Plumb_S1} phase space trajectories (panels b-e) and hovm\"oller diagrams of the mean-flow (panels f-i) for four selected values of the Reynolds number. Shown are examples for a period-1 limit cycle (panels b and f), a quasiperiodic oscillation (panels c and g), a frequency locked oscillation with frequency ratio $1/3$ (panels d and h), and a chaotic oscillation (panels e and i).

\begin{figure}
\centering\includegraphics[width=0.75\linewidth]{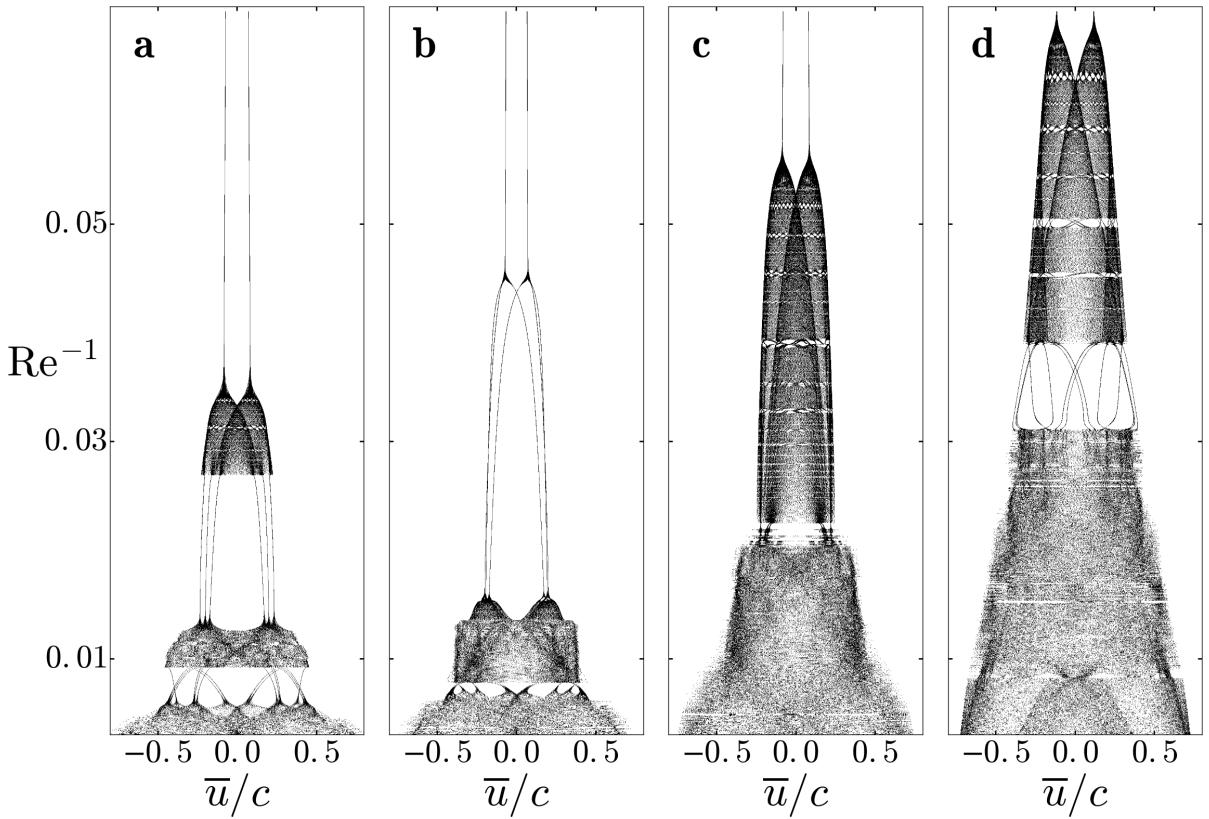}
\caption{\label{fig:BifDiag_Plumb_S2}\textbf{Additional bifurcation diagrams of the 1D quasilinear model. a.} Poincar\'{e} sections for each value of $\mathrm{Re}^{-1}$ using $\alpha=0$ (see Methods). \textbf{b.} Same for $\alpha=1/3$. \textbf{c.} Same for $\alpha=2/3$. \textbf{d.} Same for $\alpha =1$.}
\end{figure}

Additional bifurcation diagrams obtained for different values of $\alpha$ are shown in Fig. \ref{fig:BifDiag_Plumb_S2}.
Although sharing a common qualitative structure, each bifurcation diagrams show distinct interesting features. For example, Fig. \ref{fig:BifDiag_Plumb_S2}b shows that the upper quasiperiodic region vanishes almost entirely when $\alpha$ approaches $1/3$. 
Fig. \ref{fig:BifDiag_Plumb_S2}d ($\alpha=1$) shows a spontaneous breaking of the symmetry $U\leftrightarrow -U$ occurring in one of the frequency locked states ($\mathrm{Re}^{-1}\sim 0.045$), while all the frequency-locked regimes preserve this symmetry at lower values of $\alpha$ in panels a, b and c.\\

\textbf{Effect of the resolution in the quasilinear model.} In order to test the robustness of the quasilinear model results to resolution, we show in Fig. \ref{fig:BifDiag_Plumb_S4} five bifurcation diagrams for which the vertical resolution has been successively doubled from $\delta_z=3.5/15$ to $3.5/240$. Panel c corresponds to the reference resolution used in Fig. 1d. Results show a strong dependence on vertical resolution, in particular for the structure of the embedded frequency locked regimes.
However, the essential feature relevant to the periodicity disruption is the second bifurcation point $Re_{c2}$, marking the transition from periodic to quasiperiodic oscillations.
Results of Fig. \ref{fig:BifDiag_Plumb_S4} show that the value of $Re_{c2}$ is converging for a resolution $\delta_z=3.5/60$, corresponding to the reference resolution used in this work. 

\begin{figure}
\centering\includegraphics[width=1\linewidth]{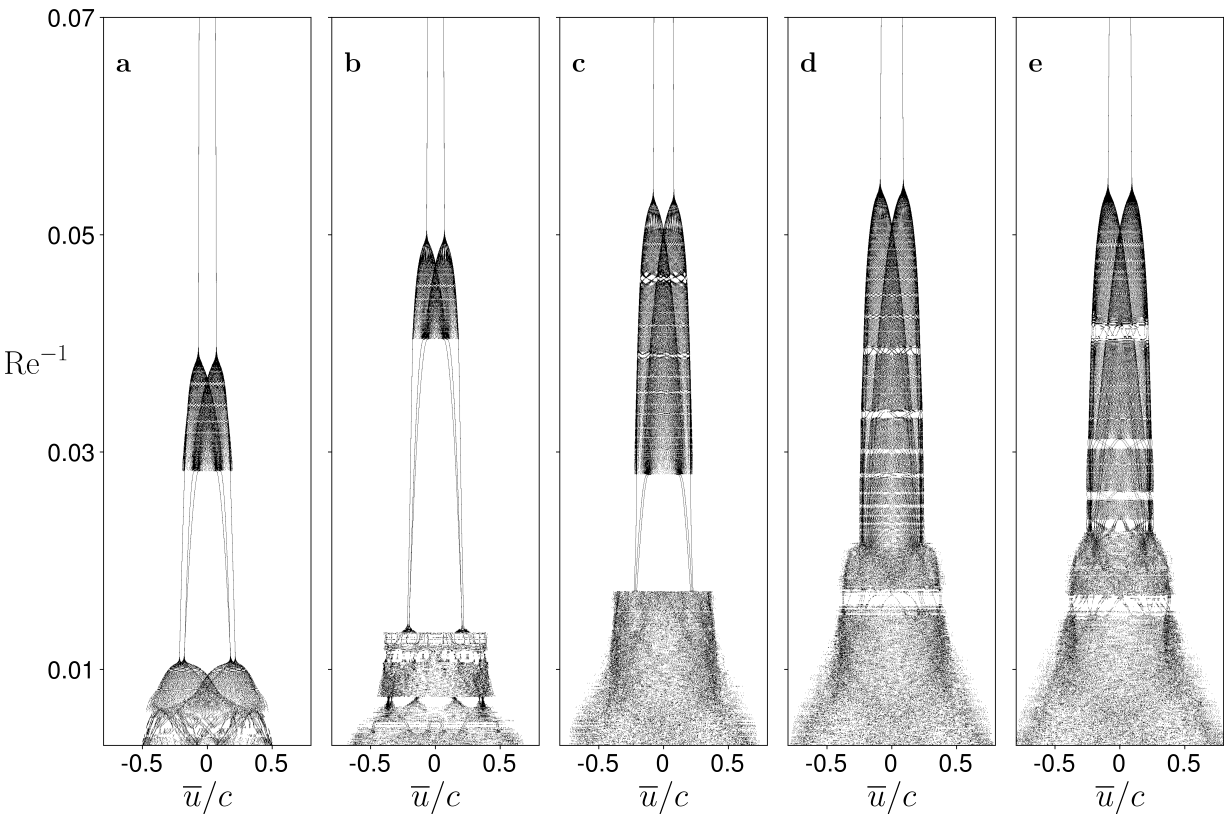}
\caption{\label{fig:BifDiag_Plumb_S4}\textbf{Resolution dependence in the 1D quasilinear model} \textbf{a} Poincar\'{e} sections for each value of $\mathrm{Re}^{-1}$ with $\alpha=0.6$. The spatial resolution used is $\delta_z=3.5/15$. \textbf{b.} Same for $\delta_z=3.5/30$.  \textbf{c.} Same for $\delta_z=3.5/60$. \textbf{d.} Same for $\delta_z=3.5/120$. \textbf{e.} Same for $\delta_z=3.5/240$.}
\end{figure}

As far as critical slowing down is concerned, the response of the system approaching the bifurcation from periodic to quasi-periodic state will be robust to higher resolutions, as the threshold $Re_{c2}$ and the nature of the bifurcation remains the same. However, the details of the response, including the vertical structure of transient oscillations and the prefactor of the recovery time power law may be affected by a change in resolution.\\

\section*{3D nonlinear simulations with rotation}

In this section, we solve the 3D Navier-Stokes equations with rotation approximated by an equatorial beta-plane. The horizontal momentum equation in (\ref{eq:NumModel}) now writes
\begin{equation}
    \begin{cases}
        \partial_{t}u+\mathbf{u}.\nabla u -\beta y v & = -\partial_{x}\phi+\nu\delta^2 u - \gamma_u \left(u-u_0\right)\\
        \partial_{t}v+\mathbf{u}.\nabla v + \beta y u &= -\partial_{y}\phi+\nu\delta^2 v
    \end{cases},\label{eq:NumModel3D}
\end{equation}
where the velocity vector field is now 3D with  $\mathbf{u}=u\hat{\mathbf{i}}+v\hat{\mathbf{j}}+w\hat{\mathbf{k}}$. $\beta$ is the Rossby parameter. Let us denote $L_{y}$ the length of the added meridional dimension and $\delta_y$ the associated resolution. We consider an horizontal aspect ratio $L_y /L_x=1$ and resolution ratio $\delta_y/\delta_x=1$, with free-slip lateral boundary condition at $y=\pm L_y /2$. We explore a weak rotation case,  for which the equatorial Radius of deformation $L_d=\sqrt{N\Lambda/\beta}$ is much larger than the meridional extension of domain: $L_d /L_y =96$. All other parameters are identical to the 2D nonlinear simulations, including the forcing, constant along the $y$ direction. The chosen initial condition breaks the meridional invariance. 

Fig. \ref{fig:Bif3D_S3} shows that bifurcation from a periodic regime (panel a) to a quasiperiodic regime (panel b) occurs when the Reynolds number is increased from $\mathrm{Re}=125$ to $\mathrm{Re}=250$. This demonstrates that the intrinsic variability observed in the 1D quasilinear model is robust to the presence of 3D wave-wave interactions.

In the Earth's stratosphere, the parameter $L_d/L_y$ is smaller than one, and equatorial waves are trapped along the equator over a typical scale of the order of the deformation radius. Future work is needed to explore the robustness of the bifurcation point $Re_{c2}$ in the presence of strong rotation.

\begin{figure}
\centering\includegraphics[width=0.5\linewidth]{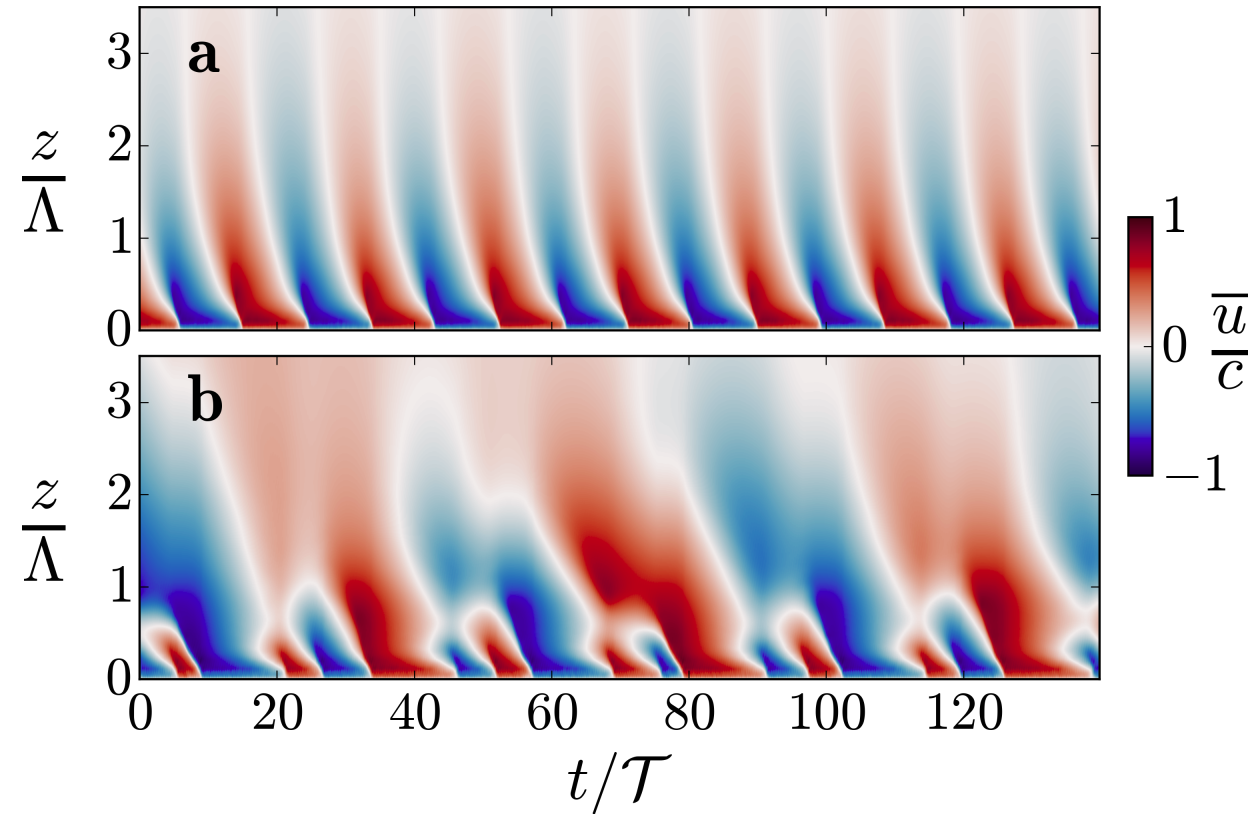}
\caption{\label{fig:Bif3D_S3}\textbf{Bifurcations in 3D nonlinear simulation with weak rotation. a.} Hovm\"{o}ller diagrams of the mean-flow $\overline{u}(z,t)$ for $\mathrm{Re}^{-1}=0.08$. \textbf{b} Same for $\mathrm{Re}^{-1}=0.04$.}
\end{figure}

\end{widetext}

\end{document}